\documentstyle[epsfig]{mn}
\newif\ifAMStwofonts
\title[The LFs of M31 and Milky Way halo clusters]{A comparison of the globular cluster luminosity functions of the
inner and outer halo of the Milky Way and M31.}

\author[JJ Kavelaars \& D.A. Hanes]
{JJ Kavelaars\thanks{e-mail:  JJ@astro.queensu.ca} and
D.A. Hanes \\
Department of Physics, Queen's University \\
Kingston,  ON  K7L 3N6 (CANADA)\\}

\date{Accepted \underline{\ \ \ \ \ \ \ \ \ \ \ \ \ \ \ }.
      Received \underline{\ \ \ \ \ \ \ \ \ \ \ \ \ \ \ };
      in original form \underline{\ \ \ \ \ \ \ \ \ \ \ \ \ \ \ }}

\pagerange{\pageref{firstpage}--\pageref{lastpage} }
\pubyear{1996}

\begin{document}

\maketitle

\label{firstpage}

\begin{abstract}
  We show that the globular cluster luminosity function (GCLF) of the
    inner halo of the Milky Way is statistically different from the GCLF
    of the outer halo. We also find a similar difference between the
inner and outer halo population of M31.
  We assert that this difference is evidence for
    some form of dynamical evolution of the cluster population and/or a
    dependence of GCLF shape on the environment in which the cluster
    population formed.  
  We also find that the turnover luminosity of the GCLF is unaffected by these
differences and further assert that this stability of the turnover luminosity
affirms its usefulness as an indicator of cosmic distance.
\end{abstract}


\section{Introduction}

  The globular cluster system of the Milky Way has long been known to be
    a multiple component system that can be divided into disk and halo
    subsystems; in addition, the halo subsystem can be further divided
    into an inner and outer halo \cite{Searl78}.  This division of the
    halo into inner and outer components has allowed the discrimination
    of the clusters along age boundaries on the basis of horizontal
    branch type \cite{Lee94}, and the picture of inside-out galaxy
    formation now seems secure. For an excellent review of these issues
    see Zinn (1996) and other works in the same volume.

  Consistent with the emerging acceptance of an inside-out picture of
    galaxy formation has come the view that the current population of
    globulars is not the primordial distribution but rather the
    surviving component of a once grander population
    \cite{Fall77,Murali96}.  Most recently, a very appealing
    picture has been presented in which outer halo clusters
    do not suffer significant disruption while those
    which are interior to  $R_{GC} \sim 8kpc$ do.
    \cite{Murali96}.

  Given both that the outer halo clusters are younger and that their
    orbits make them less likely to have undergone externally imposed
    dynamical evolution,
    the outer halo cluster population should be most like that of a
    primordial system while that of the inner halo should be evolved.
    This reasonable view can be tested by a comparison of the globular
    cluster systems (hereafter GCSs) of the inner and outer halo.  To
    allow an extension of this comparison to external galaxies, the
    comparison should be between observables which can be studied in
    remote systems.  The globular cluster luminosity function (GCLF) is
    the obvious choice.

\section{The Milky Way GCLF, inside and out}

\begin{figure}
\epsfig{file=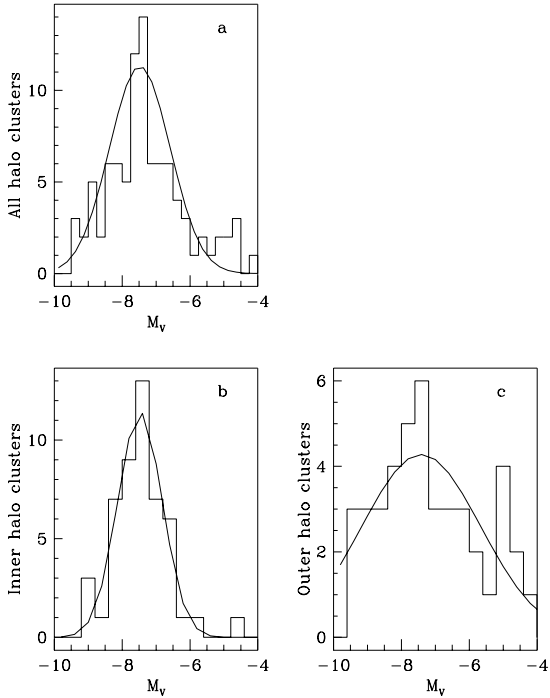,height=4in}

  \caption{(a) The luminosity function for the halo globular clusters of
    the Galaxy.  (b) The luminosity function for the inner halo
    clusters.
    (c) The luminosity function for the outer halo clusters.
    A Kolmogorov-Smirnov test shows that there is a 3.7\%
    probability that the two distributions are drawn from the same
    parent population (ie. the hypothesis that these two distributions
    are from the same parent population is rejected).  }

\label{MWgcs}
\end{figure}

  Figure~\ref{MWgcs} shows (a) the GCLF for all of the halo clusters of
    the Milky Way, and the GCLFs for the inner (b) and outer (c) halo
    subsystems.  (These data were taken from the McMaster University
    globular cluster database maintained by W.E.
    Harris\footnote{http://www.physics.mcmaster.ca/Globular.html}).
    There are approximately 50 clusters in each of the subdivisions and
    so the differences between these two distributions are decidedly
    real.  A Kolmogorov-Smirnov comparison of the two
    populations gives a probability of less
    than
    5\% that the two distributions come from the same parent population.
    This evidence strongly suggests that the inner and outer halo
    populations have different formation histories and/or different
    evolutionary histories.

  Fits of the function
  \begin{equation}
  A\exp(-(m-M_{TO})^2/(2\sigma^2))
  \label{gauss}
  \end{equation}
  to the
    three distributions are overlaid on the histograms in
    Figure~\ref{MWgcs}, with the numerical results shown in
    Table~\ref{MWfits}.  These fits were performed using the NGAUSSFIT
    routine in STSDAS under the assumption of Poisson sampling errors.
    It is of interest to note that although the inner and outer halo
    GCLFs have markedly different distributions, they have the same peak
    luminosity.  This result appears to be in conflict with recent
    dynamical models \cite{Murali96} which suggest that the peak value
    of the GCLF will become significantly fainter as the cluster
    population evolves, and that such evolution should be strongest for
    the inner halo population.

\begin{table}
 \caption{ Fits to the GCLFs for various components of the Milky
    Way GCS.
    These fits were performed using the NGAUSSFIT
 package in STSDAS.  The data are from the
 McMaster catalogue.
 Note that the peak or turnover magnitude remains constant (within
 uncertainties) for all components in the Galaxy but the outer
 halo population is significantly broader.}
\label{MWfits}
\begin{tabular}{cccc}
MW-GCS & $M_{TO}$ & $\sigma_{GCLF}$ & N \\
all clusters & $-7.44 \pm 0.15$ & $1.08 \pm 0.1 $&132 \\
all halo clusters & $-7.48 \pm 0.15 $& $0.90 \pm 0.1 $ & 93\\
inner halo clusters &$ -7.47 \pm 0.13$ & $0.66 \pm 0.1$ & 49\\
outer halo clusters & $-7.41 \pm 0.4$ & $1.76 \pm .3 $ & 44 \\
 \end{tabular}

 \end{table}

\section{The GCS of M31}
 The recognition of the two-component nature of the Milky Way halo
    cluster luminosity function immediately suggests an examination of
    the M31 GCLF under the same conditions.  To allow such a comparison 
it is necessary to divide the M31 cluster population into  disk and halo subsystems.
The  question of which M31 clusters are halo members and  which are disk 
members is confused by the lack of spectroscopic
 metallicities for a vast number of cluster candidates.   Ideally a cut of 
metallicity similar to that used for  the Milky Way would 
isolate the disk and halo
subsystems, but in practice such a straightforward approach is not possible. 
  
Previous authors have selected clusters as halo members based on their 
position on the sky \cite{Reed94}.  This selection criterion has the 
disadvantage of preferentially excluding many inner halo clusters.  
To avoid excluding objects on the basis of position
we select as our ``halo sample'' all clusters for which $(B-V) < 0.8$  in 
the Battistini {\em et al.} (1987) survey.  This selection seams justified 
in that for the Galaxy there are
no disk clusters bluer than $(B-V)=0.8$ and so this cut should eliminate the
majority of disk clusters from the M31 sample.  Unfortunately this
selection criterion also removes many true halo members from the sample;
however, we find that there is no significant difference in parameters between a
Milky Way halo sample selected as [Fe/H] $< -0.8$ and one selected using $(B-V) <0.8$.
This culling also has the advantage of 
excluding background galaxies from the Battistini {\em et al.} (1987) 
A,B sample (see \cite{Reed92} Figure 8).

\begin{figure}
\epsfig{file=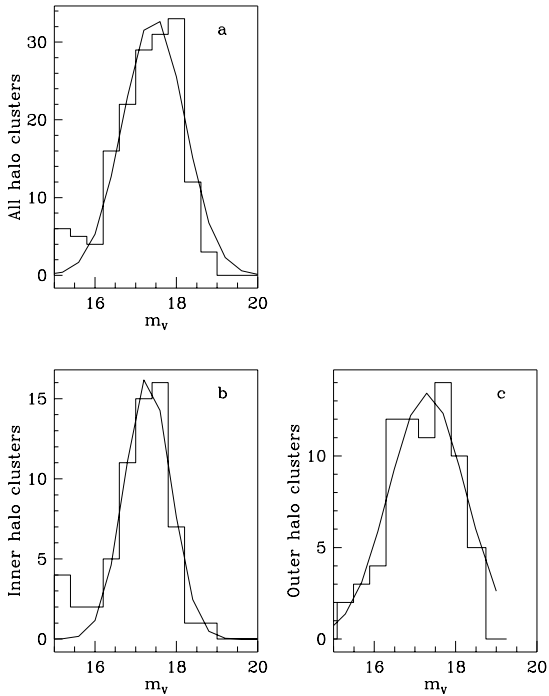,height=4in}
  \caption{(a) The luminosity function for the halo globular clusters of
    the M31.  (b) The GCLF for the inner halo clusters.  (c) The GCLF
    for the outer halo clusters.
     A Kolmogorov-Smirnov test shows that there is a 18\%
    probability that the two distributions are drawn from the same
    parent population (ie. the hypothesis that these two distributions
    are from the same parent population is not supported).  }
\label{M31gcs}
\end{figure}

  In panel (a) of
    Figure~\ref{M31gcs} we present the LF for all the globular cluster
    candidates which have (B-V) $< 0.8$ (from Battistini (1987)).
    In panel (b), we present the luminosity function for
    the subset of objects which lie within a projected radius of
    10kpc (140 arcmin for a true distance modulus of 24.45 \cite{Jacoby92}) 
	of the center
    of M31; in panel (c) we show the luminosity function for the objects
    beyond 10kpc.  This figure qualitatively reveals the same separation of the GCLF
    into peaked and flat components, when selected on the basis of
    radius, as was seen for the Galaxy.  We do not, however, consider
    this as strong a test as that afforded by the MW sample because of
    the uncertain extent to which incompleteness affects the luminosity
    functions in the subsamples at the faint end.  In addition, of
    course, there will be some contamination of the ``inner halo''
    subsample by outer halo clusters projected into the 10 kpc circle.
    Also, the determination of halo/disk membership for individual clusters
    is uncertain and it is probable that there is contamination of the
    inner halo sample by disk clusters.

  Table~\ref{M31fits} presents the results of fits of
    equation~\ref{gauss} to the various cluster populations in M31. As 
was found for the Milky Way GCS the outer halo clusters of M31 have
a broader luminosity function, and once again
there is no evidence for difference
between the turnover luminosity of the inner and outer halo.
 A K-S comparison of the data sets gives a probability of 18\% that
    the two samples are drawn from the same parent population.  This
    provides marginal
    evidence for a multiple component halo and is consistent with previous work
    \cite{Ashman93,Huchra91} which has shown that the distribution of halo
    clusters in M31 contains substructure.

  As for the Milky Way,  the
 observed difference between the two M31 halo populations cannot be the
    result of dynamical evolution of the type predicted by Murali and
    Weinberg (1996) since their model implies that the inner population
    should appear fainter than that of the outer.  There is 
    no evidence for such a shift in turnover values.

\begin{table}
 \caption{ Fits to the GCLFs for various components of the
    Andromeda galaxy (M31).
    These fits were performed using the NGAUSSFIT
 package in STSDAS.  The data are from Battistini {\em et. al}
 (1987).  Once again, the outer
 halo population is somewhat broader than that of the inner halo sub-system.}
\label{M31fits}
\begin{tabular}{cccc}
M31-GCS & $M_{TO}$ & $\sigma_{GCLF}$ & N \\
all halo clusters & $17.45\pm0.08$ & $0.76\pm0.05$ & 161 \\
inner halo clusters & $17.30\pm0.10$ & $0.57\pm0.10$ & 64\\
outer halo clusters & $17.29\pm0.17$ & $0.94\pm0.13$ & 97\\
 \end{tabular}
 \end{table}

\section{Conclusions}
 Both within the Milky Way and, to a lesser extent, M31
the outer halo clusters are clearly more indicative of a broad
luminosity function.
This population may be, then, indicative of an initial broad mass function for globular
clusters.

The inner halo clusters demonstrate a remarkably peaked distribution.
The difference between the inner and outer luminosity functions 
may indicate the fate that dynamical evolution has
in store for clusters formed near the centers of galaxies.

The turnover luminosities of the inner and the outer cluster populations
appear to be  consistent. If this is the case, then dynamical
models of cluster evolution will need to account for
the preferential stripping away of clusters fainter and brighter than 
the turnover luminosity.  This decoupling of mean luminosity from position in
the galaxies gives further assurance that GCLFs can be used as  standard candles
in cosmic distance determinations.
The dependence of the shape of the GCLF on galactocentric radius suggests that
future comparisons  should be made between cluster
populations that only include clusters of comparable radii from their host galaxy
centers.

If dynamical effects are not found to be responsible for the 
dependence of GCLF shape on radius,
 then the environment in which clusters  form is likely to be
the deciding  factor.

Many thanks to S.T. Butterworth and D. Wing for their clarifying discussion
of this issue.

\label{lastpage}

\begin{thebibliography}{}

\bibitem[\protect\citename{Ashman and Bird} 1993]{Ashman93}
Ashman, K.M., Bird, C.M. 1993 AJ 106, 2281

\bibitem[\protect\citename{Battistini {\em et. al}} 1987]{Battistini87}
Battistini, P., Bonoli, F., Braccesi, A. Federici, F. Fusi Pecci, F., Marano, B., Borngen, F., 1987, A\&AS, 67, 447

\bibitem[\protect\citename{Fall and Rees} 1977]{Fall77}
Fall, S.M., Rees, M.J. 1977, MNRAS, 181, 37P

\bibitem[\protect\citename{Huchra {\em et al.}} 1991]{Huchra91}
Huchra, J.P., Brodie, J.P., Kent, S.M., 1991, ApJ, 370, 495

\bibitem[\protect\citename{Jacoby {\em et al.}} 1992]{Jacoby92}
Jacoby, G.H. et al., 1992, PASP, 104, 599

\bibitem[\protect\citename{Lee {\em et. al}} 1994]{Lee94}
Lee, Y.-W., Demarque, P., Zinn, R.,  ApJ, 423, 248

\bibitem[\protect\citename{Murali and Weinberg} 1996]{Murali96}
Murali, C., Weinberg, M.D., 1996, MNRAS, {\em submitted}


\bibitem[\protect\citename{Reed {\em et. al}} 1994]{Reed94}
Reed, L.G., Harris, G.L.H., Harris, W.E., 1994, AJ, 107, 555

\bibitem[\protect\citename{Reed {\em et. al}} 1992]{Reed92}
Reed, L.G., Harris, G.L.H., Harris, W.E., 1994, AJ, 103, 824

\bibitem[\protect\citename{Searle and Zinn} 1978]{Searl78}
Searle, L., Zinn, R., 1978, ApJ, 225, 357

\bibitem[\protect\citename{Secker} 1992]{Secker92}
Secker, J., 1992, AJ, 104, 1472

\bibitem[\protect\citename{Zinn} 1996]{Zinn96}
Zinn, R., 1996, ed. H. Morrison and A. Sarajedini,
ASP Conf. Ser. 92, Formation of the Galactic Halo. . . .Inside and
Out, 211

\end{thebibliography}
\end{document}